# Covid-19 risk factors: Statistical learning from German healthcare claims data

Roland Jucknewitz[1], Oliver Weidinger and Anja Schramm

AOK Bayern, 93059 Regensburg, Germany

February 2021 (V2)

We analyse prior risk factors for severe, critical or fatal courses of Covid-19 based on a retrospective cohort using claims data of the AOK Bayern. As our main methodological contribution, we avoid prior grouping and pre-selection of candidate risk factors. Instead, fine-grained hierarchical information from medical classification systems for diagnoses, pharmaceuticals and procedures are used, resulting in more than 33,000 covariates. Our approach has better predictive ability than well-specified morbidity groups but does not need prior subject-matter knowledge. The methodology and estimated coefficients are made available to decision makers to prioritize protective measures towards vulnerable subpopulations and to researchers who like to adjust for a large set of confounders in studies of individual risk factors.

## 1 Introduction

In spring 2020, Covid-19 has rapidly become a public health emergency of international concern. Besides general containment approaches, such as social distancing or wearing community masks, protecting those at highest risk for severe illness is critical to prevent rising death tolls and overloaded healthcare systems. Such measures include intensified shielding, issuing of protecting masks (FFP2), and prioritization of pharmaceutical interventions during the phase of limited availability.

To be effective in these measures, there is an urgent need for a precise quantification and ranking of groups who are at increased risk of severe Covid-19. Several authors have studied the risk factors for severe outcomes like death in an overall population (Williamson et al. 2020, McKeigue et al. 2020), in the population of Covid-19-positive (Harrison et al. 2020) or in those hospitalized with Covid-19 (Karagiannidis et al. 2020). These studies assess a broad range of risk factors, especially comorbidity conditions, and some also consider risk prediction to rank the most vulnerable individuals (Clift et al. 2020). Beyond these, numerous studies focus on single risk factors such as diabetes (Barron et al. 2020) or HIV infection (Bhaskaran et al. 2020).

Wherever they are examined in the Covid-19 literature, (co-)morbidity risk factors are identified by a few broad risk groups such as coronary heart disease or diabetes mellitus. Usually, these groups, whose number is typically in the order of a dozen and effectively range to 38 (Clift et al. 2020), are specified a priori, e.g., by referring to risk groups from other infectious respiratory diseases like influenza, or from groups defined for other purposes, such as the Charlson or Elixhauser comorbidity groups or the corresponding indices.

In this study, we try to identify risk factors from a German population of a statutory health insurance without pre-specifying such groups. The risk index, which can be constructed from the factors, allows the public health system to prioritize protective measures, while research studying single risk factors or a limited set thereof can use it to statistically account for the most relevant comorbidities within a single variable.

---

[1] Corresponding author. Email: roland.jucknewitz@by.aok.de. Phone +49 941 79606 309.

# 2 Methods

We use data from AOK Bayern, a regional statutory health insurance with a market share of approximately 40% in the German region of Bavaria. Our population are all persons insured by AOK Bayern on 1/3/2020 (N=4,636,379). We exclude individuals who finished insurance status before 1/12/2020, persons who were not enrolled during the entire year 2019 and non-German residents (identified by postal address or by the AUS-AGG label used in the risk adjustment scheme, RSA, among the German statutory health insurance funds). Thus, our population-at-study includes 4,101,745 individuals. The population is not restricted to Covid-19-positive or hospitalized patients, so that the analysis is designed to measure an overall population risk that includes both the infection risk and the risk given an infection. For this population we collect predictors and outcomes in an anonymized dataset from different healthcare sectors such as ambulatory or hospital treatments and drug prescriptions as well as socio-economic factors - all of them available at high data quality.

We treat severe, critical and fatal clinical courses of Covid-19 as binary outcomes. A severe Covid-19 case (Y1) is defined by an admission to hospital with confirmed Covid-19 diagnosis (ICD[2]U07.1) and either documented pneumonia (ICD J12.- to J18.-) or mechanical ventilation. A critical course (Y2) restricts this set to persons who received intensive care or mechanical ventilation, who had a documented sepsis (ICD R65.0, R65.1 or R57.2) or who died within hospital (identified by reason of discharge). Intensive care is identified if the first, last or longest station at hospital is an intensive care unit (ICU), at least one diagnosis is coded by an ICU or by the OPS keys 8-980.-, 8-98D.-, 8-98F.-, 8-712.0, 8-721.1, 8-721.2 or 8-721.3. In-hospital-death (Y3) further restricts this set to those who died within hospital. We consider critical course (Y2) as primary outcome, while we use Y1 and Y3 in additional analyses to assess the sensitivity with respect to outcome. Outcome data are collected from all hospital admissions on 1/3/2020 or later on for which final invoices were available on 22/11/2020. We use data that become available later on until 25/1/2021 for a prediction validation set in section 4.

To identify risk factors we use the following predictors: age and gender in 40 groups (age is categorized by 5%-percentile), nationality (the 10 most frequent nationalities and "other"), nursing home residence, insurance and occupational status (compulsorily insured employee, voluntarily insured employee, compulsorily insured pensioners, voluntarily insured pensioners, roughly short-term unemployed ("Arbeitslosengeld 1"), roughly long-term unemployed ("Arbeitslosengeld 2"), welfare recipient, self-employed, dependently insured spouse, dependently insured child), and labour income by quintile groups. We approximate the regional infection risk by the 7-day Covid-19-incidence from Robert Koch Institute, measured on the date of admission to hospital for individuals with the considered outcome. For all other individuals, the reference date is randomly assigned from a distribution that matches the distribution of the admissions to hospital for the aforementioned Covid-19 cases. Hence, regional biases due to regional disparities of infection rates and morbidities are mitigated. Whenever nationality is missing (N=27), we set it to German as the most probable outcome. No other characteristics are missing in our dataset.

With regard to the risk that is due to morbidity and medical treatments, we use data on outpatient and inpatient diagnoses according to the ICD, prescriptions of drugs from ready-made medicines or mixtures according to the German Anatomical Therapeutic Chemical (ATC)-Classification, as well as medical interventions according to the German procedural classification (Operationen- und Prozedurenschlüssel or OPS); for information about the classification systems see BfArM[3]. Data cover the period from 1/1/2019 until 29/2/2020. As an exception, ambulatory diagnoses are only used until 31/12/2019, since these data merely refer to a quarter as a whole.

---

[2] International Statistical Classification of Diseases, German Modification or ICD-10-GM
[3] https://www.dimdi.de/dynamic/en/classifications/

Our aim is to avoid a loss of medical information by prior grouping, and nonetheless to structure the information where it is found appropriate. We therefore retain data on the granular level of full digit ICD and ATC, and 5-digit OPS codes, and add the hierarchical structure of the classification systems of ICD, ATC and OPS into the information set (see Stausberg 2015 who suggests using ICD structure on chapter and group levels). Hence, if a high level code such as the 5-digit-ICD I25.22 ("Old myocardial infarction, more than one year past", hierarchical level 5) is coded, we activate an according binary variable, but also assign dummies for the lower levels of the hierarchy, i.e., the 4-digit-code I25.2 "Old myocardial infarction" (level 4), the 3-digit-code I25 "Chronic ischaemic heart disease" (level 3), the group I20-I25 "Ischaemic heart diseases" (level 2) and the chapter IX "Diseases of the circulatory system" (level 1). The same logic holds for ATC codes, where anatomical main group, therapeutic, pharmacological and chemical subgroup as well as chemical substance are the five hierarchical levels. For OPS keys, where groups, 3-, 4- and 5-digit codes are the second to fifth level analogously to ICD, we use only codes from the chapters 5 (operations), 6 (medications) and 8 (non-operative therapeutic interventions) and drop the chapter code. This total enriched set of medical information amounts to a number of 33,362 binary predictor variables, from which risk factors are to be identified; see table 1.

We use a logistic lasso approach to predict the binary outcomes by a selected set of predictors and shrinkage, based on the glmnet 4.0 package in R, see Friedman et al. (2010). To favour the selection on low-level rather than high-level codes where appropriate we apply a differential shrinkage factor that corresponds to the hierarchical level of the medical codes, while first level codes are shrinked with the same factor of 1 as demographic and socio-economic variables. No shrinkage is applied to the regional incidence, since this variable is measured at a different (non-binary) scale than all other variables and bias is avoided in this variable. 5-fold cross validation is applied to maximize the area under the ROC curve (AUC or C-statistic) with respect to the shrinkage parameter. For each medical category (or code) we obtain its estimated contribution to Covid-19 risk at logistic scale by adding the corresponding parameters of all hierarchical levels.

To assess the predictive performance of the chosen approach, we compare the described method of full information to prior approaches of pre-specified Covid-19 risk groups. To be precise, we reconstruct medical groups of Williamson et al. (2020), of McKeigue et al. (2020), Schröder et al. (2020) and Clift et al. (2020), and also apply the Charlson Morbidity groups used by Harrison et al. (2020). Furthermore, we quantify the possible improvement that nonlinear machine learning techniques can have over the linear lasso, and add tree boosting with full quantitative information to the comparison. As a first evaluation, the predictive performance is assessed in the same cross validation folds as mentioned above, so that variable selection and parameter estimation does not use the fold to be predicted beyond the univariate maximization of the shrinkage parameter. A second evaluation setup is constructed from outcome data that became available after all model specification and estimation steps where finished. We compare the AUC, the expected information for discrimination $\Lambda$, and predictive likelihoods for statistical comparisons.

We construct a risk index which is suitable to adjust for comorbidities when investigating the effect of specific variables in greater detail. To this end, we compute the logistic link value on the cross-validation folds. No outcome data from the same folds are used to compute the index, and hence it can be consistently used on the right side of a regression explaining the outcome. The contributions of risk factors to be investigated in greater detail can be easily cancelled by setting their coefficients in the lasso model of each fold to zero.

**Table 1**: Number of codes per hierarchic level.

| Level | ATC | ICD | OPS |
|---|---|---|---|
| 1 | 14 | 22 | (dropped) |
| 2 | 99 | 241 | 43 |
| 3 | 275 | 1,697 | 137 |
| 4 | 1,023 | 8,876 | 953 |
| 5 | 6,787 | 5,514 | 7,681 |

## 3 Estimation results

An overview over population characteristics by outcome groups is provided in table 2. Of the 4,101,745 individuals in the study population, 1,611 are characterized by the primary outcome critical Covid-19 (Y2), with admissions to hospital occurring between 1/3/2020 and 10/11/2020, while the broader outcome (Y1) is documented for 3,731 and the hospital mortality outcome (Y3) is documented for 931 individuals. Age, gender, multimorbidity (measured as number of distinct ICD groups) and polymedication (number of distinct pharmaceutical subgroups) differ substantially between the groups, with the more severe outcomes dominated by a greater percentage of male, older and more multimorbid and polymedicated patients.

Estimating the baseline logistic lasso regression for the primary outcome Y2 over the full population, we obtain an optimal shrinkage parameter $\lambda = 3.4882 \times 10^{-7}$ that yields an in-sample AUC of 0.890. All coefficients are available online as supplementary table S1. In this setup, 4,497 coefficients are nonzero, among them 3,292 ICD, 792 ATC and 413 OPS codes. While all of the 1$^{st}$ level (e.g. ICD chapters) and 85.5% of 2$^{nd}$ level codes (e.g. ICD groups) are nonzero, only 10.5% of 5$^{th}$-level codes differ from zero. This shows the necessity to subtly differentiate some, but not all of the major morbidity groups. Due to the hierarchical structure, each ICD and ATC code has a nonzero coefficient when aggregated over hierarchical levels.

The most severe single risk factors (by magnitude of the odds ratios and applying a minimum group size of 1,000) are nursing home (OR 3.0), three diuretics in the ATC class C03CA (Torasemid, OR 2.3) and the ICD I98.2 (Oesophageal varices without bleeding in diseases classified elsewhere, OR 2.2). Overall, diagnoses related to the circulatory system and pharmaceuticals for the cardiovascular system are dominant in the medical factors with highest impact and account for 97 of the 100 top codes. Categorial groups of age and gender are not leading this list, although the effects are interdependent with socio-economic characteristics like pension status.

**Table 2**: Descriptive statistics of selected independent variables, by outcome status.

|  | No outcome | Y1 severe | Y2 critical | Y3 death |
|---|---|---|---|---|
| N | 4,098,014 | 3,731 | 1,611 | 931 |
| Age | 45.5 | 69.9 | 73.9 | 79.8 |
| Gender (male) | 48.4% | 54.8% | 59.2% | 56.1% |
| Multimorbidity | 11.3 | 22.2 | 24.0 | 26.5 |
| Polymedication | 3.5 | 7.7 | 8.5 | 9.4 |
| Nursing home | 1.1% | 13.7% | 17.3% | 25.1% |
| Regional incidence | 56.8 | 81.2 | 74.4 | 82.3 |
| Nationality: German | 81.0% | 81.5% | 86.5% | 93.1% |
| Nationality: Other | 19.0% | 18.5% | 13.5% | 6.9% |
| Wage: na/not working | 57.6% | 79.6% | 87.7% | 96.9% |
| Wage: 1th quintile | 8.5% | 2.9% | 1.6% | 0.9% |
| Wage: 2nd quintile | 8.5% | 4.5% | 2.4% | 0.4% |
| Wage: 3rd quintile | 8.5% | 4.8% | 2.9% | 0.3% |
| Wage: 4th quintile | 8.5% | 3.7% | 2.0% | 0.4% |
| Wage: 5th quintile | 8.5% | 4.5% | 3.4% | 1.1% |

Considering aggregated (2nd level) groups only, we obtain aggregated log odds ratios (logOR) by averaging over all fine-grained codes and weighting by their prevalence. The importance of the group in the population (Pop) is obtained by multiplying the log odds ratios by group size and ranking over this excess relative risk. The top 20 groups with respect to the critical outcome Y2 are shown in table 3. Hypertensive diseases (ICD I10-I15), other forms of heart disease (ICDs I30-I52) and agents acting on the renin-angiotensin system (ATC C09) add most to the population-level risk, followed by metabolic disorders (E70-E90), which is the first group not related to the circulary system.

Due to the ceteris paribus notion of (logistic) regression coefficients, identifying a risk factor as having a relatively high coefficient does not mean that the corresponding individuals are also at highest risk from a ranking point of view. The overall effect of having a certain health condition is substantially affected by complementary medication and by its co-morbidities. Hence, while the regression-based odds ratio of the oldest male age group (AGG20) is higher than the odds ratio for end-stage kidney disease (ICD N18.5), the morbidity risk including possible further conditions like dialysis of the latter group dominates that of AGG20 (OR 17.6 versus 4.6), and also the overall risk (including the age effects) of individuals with end-stage kidney disease exceeds that of AGG20. We therefore emphasize that a ranking of risks should be in an additive manner accounting for all risk factors of the individuals rather than looking at the risk factors with the highest odds ratios alone.

**Table 3**: Adjusted log odds ratios (logOR) and ranks in population importance (Pop) of aggregated medical risk factors at aggregation level 2. The population importance is derived by multiplying the log odds ratios by group size.

| | **Outcome** | **Y1 severe** | | **Y2 critical** | | **Y3 death** | |
|---|---|---|---|---|---|---|---|
| **Code** | **Name** | logOR | Pop | logOR | Pop | logOR | Pop |
| I10-I15 | Hypertensive diseases | 0,307 | 1 | 0,385 | 1 | 0,350 | 1 |
| I30-I52 | Other forms of heart disease | 0,319 | 3 | 0,440 | 2 | 0,430 | 3 |
| C09 | Agents acting on the renin-angiotensin system | 0,363 | 4 | 0,511 | 3 | 0,538 | 2 |
| E70-E90 | Metabolic disorders | 0,263 | 2 | 0,261 | 4 | 0,135 | 8 |
| C03 | Diuretics | 0,561 | 7 | 0,787 | 5 | 0,867 | 4 |
| E10-E14 | Diabetes mellitus | 0,299 | 5 | 0,299 | 6 | 0,177 | 14 |
| C07 | Beta blocking agents | 0,339 | 13 | 0,522 | 7 | 0,556 | 5 |
| I20-I25 | Ischaemic heart diseases | 0,307 | 10 | 0,421 | 8 | 0,369 | 9 |
| A02 | Drugs for acid related disorders | 0,280 | 12 | 0,377 | 9 | 0,351 | 7 |
| C10 | Lipid modifying agents | 0,316 | 14 | 0,496 | 10 | 0,536 | 6 |
| I80-I89 | Diseases of veins, lymphatic vessels and lymph nodes, not elsewhere classified | 0,317 | 11 | 0,400 | 11 | 0,329 | 12 |
| A10 | Drugs used in diabetes | 0,287 | 16 | 0,418 | 12 | 0,399 | 11 |
| E00-E07 | Disorders of thyroid gland | 0,245 | 6 | 0,194 | 13 | 0,092 | 24 |
| E65-E68 | Obesity and other hyperalimentation | 0,321 | 8 | 0,282 | 14 | 0,108 | 27 |
| C08 | Calcium channel blockers | 0,343 | 24 | 0,516 | 15 | 0,529 | 15 |
| I70-I79 | Diseases of arteries, arterioles and capillaries | 0,289 | 22 | 0,430 | 16 | 0,414 | 18 |
| I60-I69 | Cerebrovascular diseases | 0,327 | 20 | 0,427 | 17 | 0,332 | 20 |
| B01 | Antithrombotic agents | 0,231 | 25 | 0,240 | 18 | 0,410 | 13 |
| N17-N19 | Renal failure | 0,245 | 27 | 0,274 | 19 | 0,426 | 17 |
| N02 | Analgesics | 0,161 | 15 | 0,115 | 20 | 0,221 | 10 |

The effects are strongly correlated between the outcomes, as can be seen in the top groups in table 3, and in supplementary figure S1 which shows the scatter plot of log odds ratios corresponding to each predictor. The overall risk score on logistic scale follows a right-skewed distribution, see supplementary figure S2. From the distribution of scores over age groups younger and older than 80, it becomes evident that even if age is a significant risk factor, morbidity-based criteria often outweigh a higher age, and 41% of the top-5%-scorers (with score at probability scale greater than 0.00151) are neither 80 years old nor reside in a nursing home.

We illustrate the use of our risk index to adjust for a large set of confounders. To this end, we focus on the risk effect of age and gender. To construct the index, we compute predictions of the baseline model at logistic scale, and – in concordance to the question at hand – leave out age and gender groups in the predictions by setting their coefficients to zero. Using this index on the right hand side of a regression that explains the outcome would result in a bias, since the same outcome enters also on the right hand side of the regression through the coefficients of the index. Therefore, cross validation is applied so that for each data fold the model to construct the index is estimated only from data outside this fold.

To study age and gender effects we estimate a generalized additive model where age effects are modelled by smooth cubic splines separated by gender, and the risk index enters the model linearly. Age and gender effects from the full model are compared to unconditional age and gender effects without adjustment for other confounders such as comorbidities. To keep the effects comparable, we account for socio-demographic groups of pensioners and co-insured children even in the unconditional model. The risk profiles are shown in figure 1. It is evident that the rise in risk that is due to older age and being male is highly overstated by the unconditional profile in the lower panel, while the sharp rise from age 65 on is not visible in the upper panel, where the comorbidity risks are held constant. Still, age and gender remain statistically highly significant and are – considered individually – among the most important risk factor even in the conditional model. Other risk factors or morbidities can be studied similarly, by cancelling related variables when computing the risk score and running individual regression with a more detailed specification of the risk factors at question.

**Figure 1**: Risk structure of age and gender conditional on risk index (upper panel) and unconditional (lower panel). For the upper panel, the risk index at logistic scale is set to its mean for all individuals. Both panels are conditional on the strongly age-dependent social status as co-insured child or pensioner to inforce comparability.

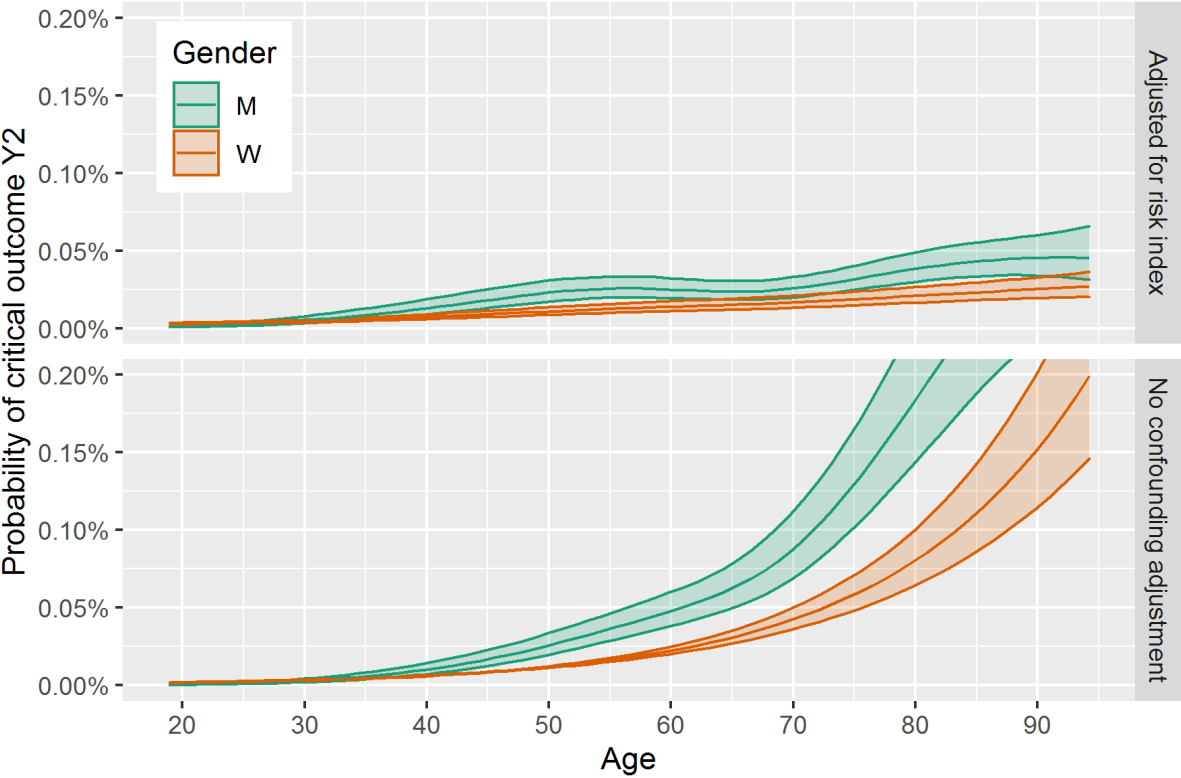

## 4 Predictive Performance

We evaluate the predictive performance of the described high-dimensional regression approach and compare it to several benchmarks. To that end, we obtain predictions on 5-fold cross validated samples, where outcome data of the same fold have neither been used to select the variables nor to compute the predictions. Additionally, we assess the predictive performance in an a priori unseen holdout sample of recent data which became available after the models have been estimated.

Firstly, we assess whether there are gains in using widely unstructured medical data in our baseline model as compared to using age and gender alone. Additionally, we use only age, gender and the described socio-demographic variables together with risk groups defined in the prior literature. We adjust all predictions at logistic scale to fit the overall prevalence of the outcome, since with one method (boosting, see below) the scale is shifted by differential weights on outcomes and controls.

The results of this predictive comparison can be seen in table 4. Consider the middle panel corresponding to the primary outcome Y2 and the left three numeric columns corresponding to cross validation first. All three measures lead to the same ranking of models, where the baseline model outperforms the group-based approaches. While the area under the curve (AUC, aka C-statistic) is a widely used metric, we compute also the expected weight of evidence $\Lambda$ promoted by McKeigue (2019) which was used in McKeigue et al. (2020). A doubling of $\Lambda$ ("adding one bit of information") is understood as the quantity of information that halves the hypothesis space. Hence, adding socio-demographic and medical information in the baseline model adds a substantial amount of information to the age and gender model. The predictive log likelihoods show a high statistical evidence in favour of the baseline model as compared to the smaller group-based models. E.g., the log likelihood difference of the baseline and the McKeigue et al. (2020) model is about 160. It is notable, however, that the groups by McKeigue et al. (2020) and Williamson et al. (2020) outperform the baseline model in the cross validation framework for outcome Y1, and also for some performance measures with outcome Y3, see the upper left and lower left panels of table 4.

In the last line of each panel, we add tree boosting as a nonlinear approach which is considered the state of the art in machine learning using tabular data. We use the XGBoost library, version 1.0.0.2, in R, where tree depth is set to 3, the shrinkage parameter is 0.01, the number of iterations is determined by cross validation and the scale_pos_weight parameter is used to weight the outcome group equally to the much larger group of controls in the overall population. Each panel of table 4 shows that the nonlinear approach improves over the baseline model with a slightly larger AUC and a statistically relevant log likelihood difference (of 89 in the middle left panel). The ROC curve in figure 2 shows that the baseline model performs similarly to boosting for the highest-risk groups at the left, while persons with low-risk profile are ranked slightly worse. The steep slope for the highest-risk groups at the left illustrates the potential to avoid a high proportion of critical outcomes by properly prioritized protective measures of only a small number of effectively protected individuals.

A second evaluation of predictive accuracy uses only data on outcomes that have become available after the estimation of the models and consists of all hospital stays for which invoices have entered the database between 23/11/2020 and 25/01/2021. We thus assess the stability of the models over time and

**Table 4:** Predictive performance measures of different models based on cross validation setup (left three columns) and test sample of newly available outcome data (right three columns) for different outcomes (upper, middle and lower panel).

| Model | Setup 1: Cross validation | | | Setup 2: New data | | |
|---|---|---|---|---|---|---|
| | AUC | Lambda | logLik | AUC | Lambda | logLik |
| **Outcome Y1** | | | | | | |
| Baseline | 0.839 | 0.783 | -26,442 | 0.846 | 0.754 | -33,540 |
| Age and gender | 0.798 | 0.222 | -28,102 | 0.817 | 0.221 | -35,520 |
| McKeigue et al. (2020) | 0.842 | 0.945 | -26,380 | 0.833 | 0.630 | -33,928 |
| Williamson et al. (2020) | 0.844 | 0.940 | -26,335 | 0.833 | 0.599 | -33,956 |
| Schröder et al. (2020) | 0.835 | 0.912 | -26,531 | 0.825 | 0.555 | -34,203 |
| Charlson groups | 0.837 | 0.916 | -26,481 | 0.826 | 0.531 | -34,169 |
| Clift et al. (2020) | 0.805 | 0.740 | -27,126 | 0.828 | 0.541 | -34,096 |
| Tree boosting | **0.855** | **1.103** | **-26,074** | **0.861** | **1.123** | **-33,160** |
| **Outcome Y2** | | | | | | |
| Baseline | 0.875 | 1.067 | -12,277 | 0.888 | 1.047 | -14,562 |
| Age and gender | 0.843 | 0.270 | -13,194 | 0.868 | 0.268 | -15,658 |
| McKeigue et al. (2020) | 0.865 | 0.829 | -12,437 | 0.878 | 0.821 | -14,770 |
| Williamson et al. (2020) | 0.863 | 0.796 | -12,438 | 0.880 | 0.790 | -14,759 |
| Schröder et al. (2020) | 0.856 | 0.755 | -12,528 | 0.871 | 0.749 | -14,904 |
| Charlson groups | 0.857 | 0.731 | -12,516 | 0.873 | 0.725 | -14,873 |
| Clift et al. (2020) | 0.858 | 0.744 | -12,483 | 0.874 | 0.737 | -14,828 |
| Tree boosting | **0.888** | **1.365** | **-12,181** | **0.900** | **1.453** | **-14,440** |
| **Outcome Y3** | | | | | | |
| Baseline | 0.935 | 1.593 | -7,027 | 0.937 | 1.552 | -9,516 |
| Age and gender | 0.906 | 0.510 | -7,700 | 0.918 | 0.511 | -10,339 |
| McKeigue et al. (2020) | 0.932 | 1.866 | -7,032 | 0.930 | 1.246 | -9,663 |
| Williamson et al. (2020) | 0.933 | 1.815 | -7,013 | 0.931 | 1.193 | -9,681 |
| Schröder et al. (2020) | 0.930 | 1.767 | -7,060 | 0.925 | 1.149 | -9,790 |
| Charlson groups | 0.931 | 1.760 | -7,036 | 0.928 | 1.120 | -9,742 |
| Clift et al. (2020) | 0.927 | 1.136 | -7,144 | 0.929 | 1.123 | -9,709 |
| Tree boosting | **0.937** | **2.059** | **-6,998** | **0.944** | **2.081** | **-9,425** |

**Note**: The measures are area under the curve (AUC), the expected weight of evidence (Lambda) and the predictive binomial log likelihood (logLik) and are shown for all three outcomes considered in the paper. Lambda is computed as the mean of $w_i = (2y_i - 1)[\log(p_i(1-p_i)^{-1}) - \log(p(1-p)^{-1})]$, where $y_i$ is the outcome, $p_i$ is the predicted probability and $p$ is the prior probability (we use the overall prevalence of the outcome, which is 0.000393 for outcome Y2).

with respect to the larger time lag between outcomes and input data (where still only data until 28/2/2020 are used). We estimate the models with the whole population and with outcome data available until 22/11/2020. The test set consists of all individuals who have not incurred a considered outcome in the previously available dataset. The results in the three right columns of table 4 show that the ranking of methods are similar to that of the cross validation study. The baseline model is consistently better than group-based approaches, however, while the superiority of the tree boosting approach is somewhat more pronounced.

**Figure 2**: Receiver operating characteristic (ROC) curve from three predictive models based on cross validation sets.

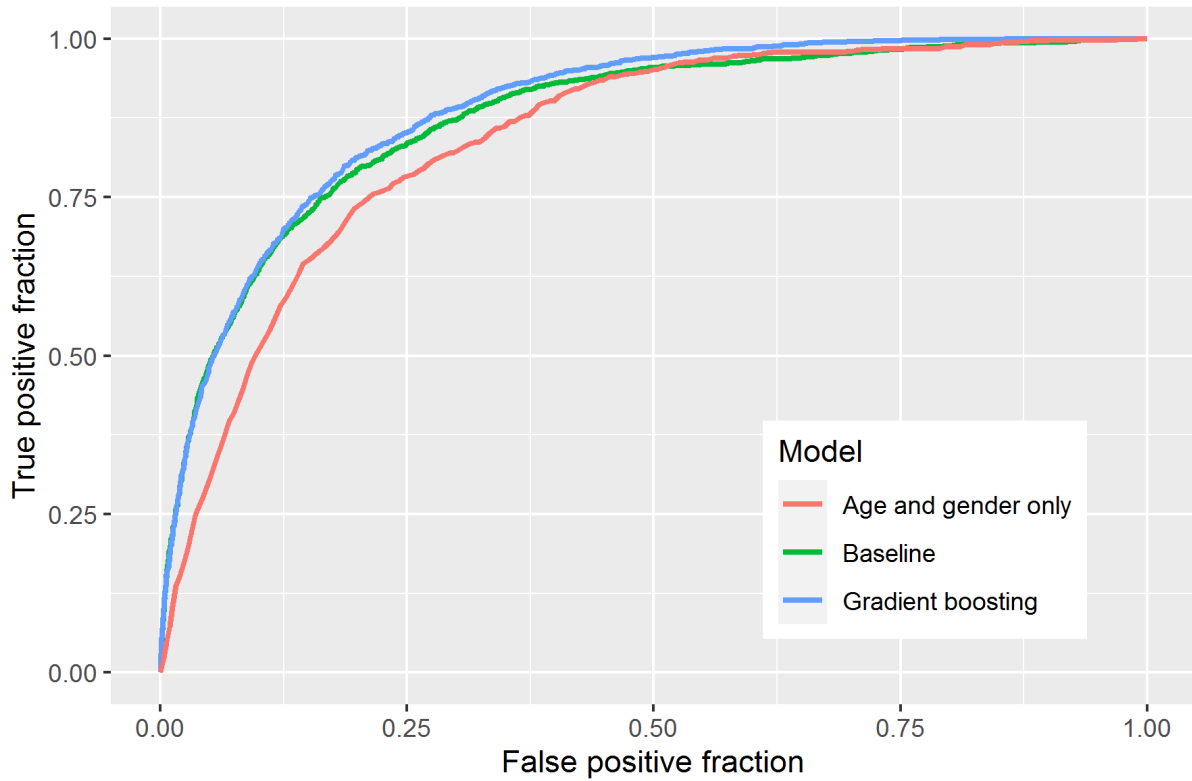

We conclude that fine-grained medical information is able to improve predictions of critical Covid-19 outcomes in an overall population, while using predetermined risk potentially misses important information and can thus have an adverse effect on prediction performance.

## 5 Discussion

We developed a statistical learning approach which is successful in identifying risk factors and competitive with respect to predictive performance. However, several issues need to be addressed to properly interpret the results of this study.

Firstly, we have studied risk in a population of AOK Bayern insured. The population, while composed from nearly all groups of the Bavarian population (besides civil servants and the small group that is not health insured) is not representative for the overall population due to differential selection of public and private health insurers.

Secondly, studying risk in an overall population might not give appropriate answers to the question who is at highest risk *when infected* with Covid-19, as shortly stated in the introduction. Differences between the results of these two concepts occur when considered risk groups have different risks of *getting infected* because of e.g. different protective measures between these groups. Another possible source of bias is heterogeneous regional Covid-incidence when also the risk factors are not evenly dispersed regionally. We tried to account for this by including the regional incidence as a predictor.

Thirdly, the risk groups and coefficients identified by the statistical learning procedures are not always easy to interpret from a medical point of view. One reason is that diagnoses and treatments (drugs, operations) are introduced as separate variables. Their effect might be cumulated (e.g. adding the effects of diabetes diagnoses and of antidiabetic drugs), but in many cases the treatments are not all specific for certain medical conditions. Thus, when studying medical risk groups in greater detail, it might be necessary to build the corresponding risk variables from a subject-specific perspective, while other confounding risks can be accounted for using our approach as outlined in the discussion of age and gender risks above.

Fourthly, we have used country-specific medical codes, and the risk coefficients might not be representative beyond the coding environment of the German statutory health insurance system. For other setups, it might be necessary to use other coding systems and a re-estimation of the coefficients, but the overall methodology should still be applicable.

Finally, the aim of the study was to consider fine-grained medical information and to assess their informative content, but the baseline model studied here is not designed to reach maximum predictive performance, as results with the gradient boosting benchmark show. Superior predictive performance might be obtained by combining risk groups and fine-grained information in a nonlinear machine learning algorithm. For practical purposes (e.g. prioritization), however, predictive performance might not be the only target. Ethical issues (leaving out nationality or other socio-economic factors) or arguments for a simpler more practicable model (grouping on lower levels and linear score models) could outweigh small boosts in predictive accuracy in most practical situations.

# 6 Conclusion

We have identified and quantified risk factors of critical Covid-19 outcomes in the overall population of AOK Bayern insured. Hypertensive and other heart diseases and their medication appear most relevant in the population. The methodological approach of using fine-grained hierarchical information from medical classification systems succeeds in better predictions of the outcomes than the use of pre-specified morbidity groups – in particular no subject-matter knowledge is needed for our baseline model. The methodology and our published coefficients may be of interest for researchers aiming at confounder adjustment in their study of individual risk factors even for smaller study cohorts. If predictive performance is the primary goal, a nonlinear approach like tree boosting that also uses fine-grained information seems particularly promising.

# Funding information

The authors received no external funding. They have no conflicts of interest to declare.

# Ethics statement

The data used in this analysis were completely anonymized. Therefore, an approval of an ethics committee or another respective institution was not required. Because of anonymization, a written consent was not required as well.

# Supplementary Tables

**Table S1**: Coefficients of the baseline model. It is available as a separate file from https://github.com/Rolehand/LearnCovid19Risk/blob/main/JucWeiSc2021_TableS1.xlsx

# Supplementary Figures

**Figure S1**: Scatter plot of log odds ratios corresponding to each ICD-, ATC and OPS-code-based predictor from baseline model estimated for different outcomes Y1 and Y2.

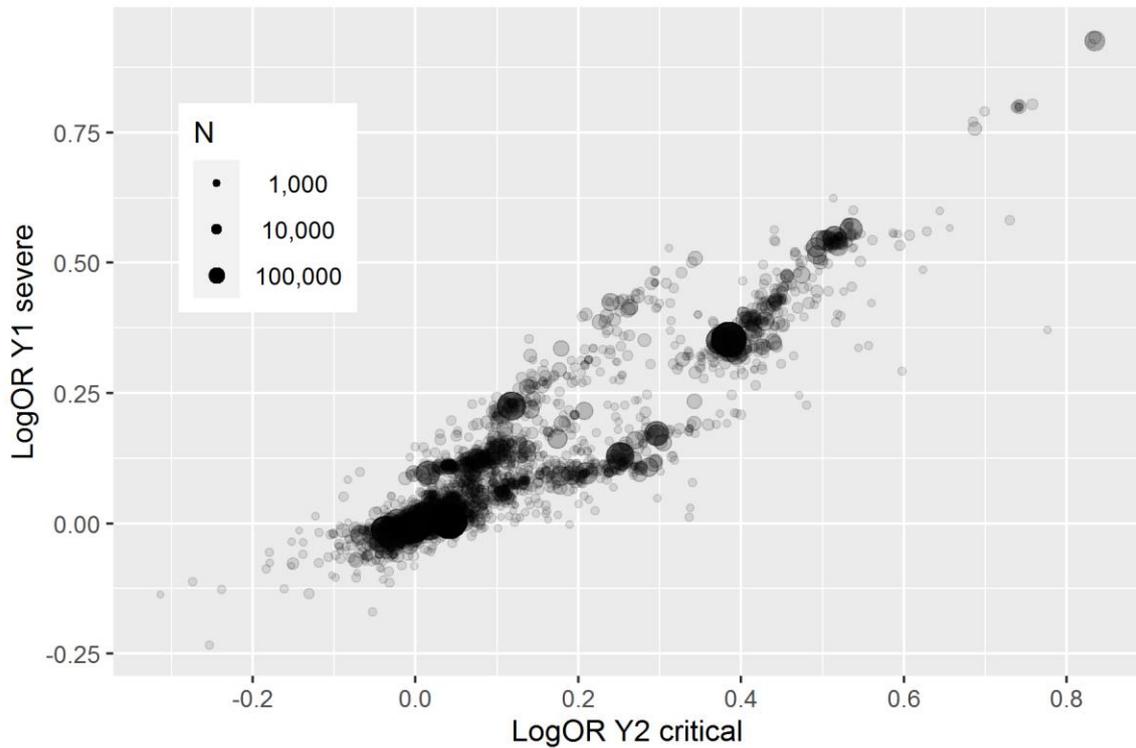

**Figure S2**: Histogram of predictions from baseline model (risk score) on logistic scale, by age groups.

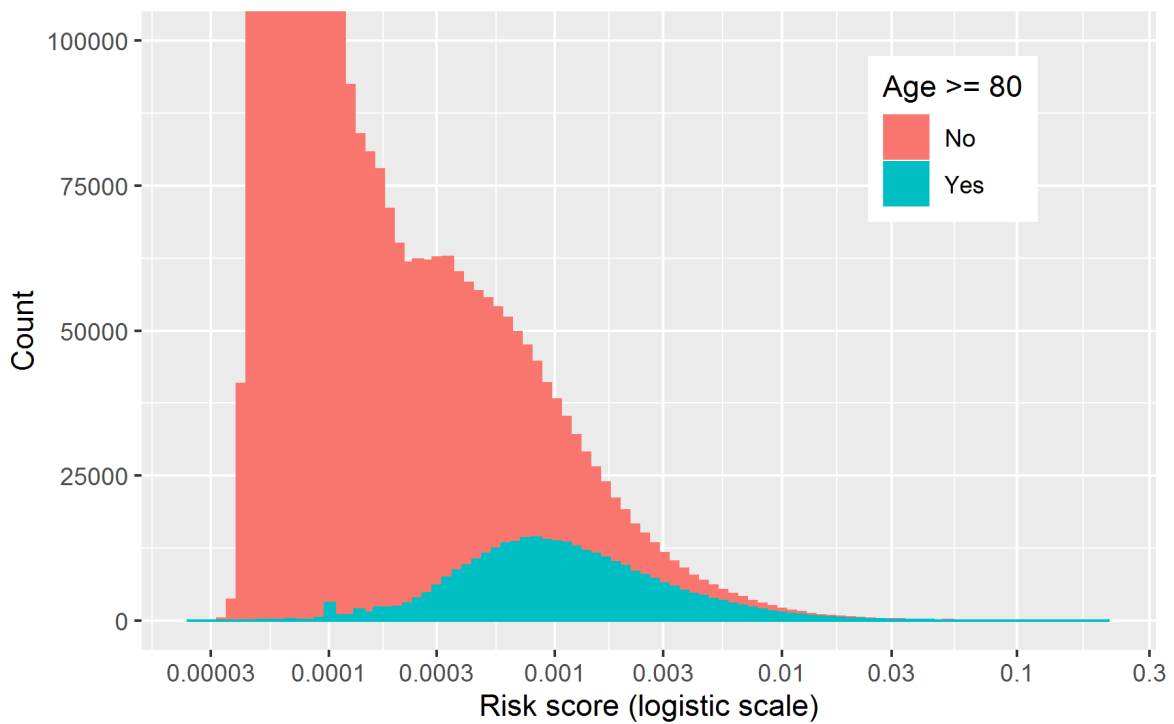